# A silicon-based microelectrode array with a microdrive for monitoring brainstem regions of freely moving rats

**G Márton[1,2], P Baracskay[3], B Cseri[3], B Plósz[4], G Juhász[3], Z Fekete[2,5], A Pongrácz[2,5]**

1       Comparative Psychophysiology Department, Institute of Cognitive Neuroscience and Physiology, Research Centre for Natural Sciences, Hungarian Academy of Sciences, 2 Magyar Tudósok Blvd., H-1117, Budapest, Hungary

2       MEMS Laboratory, Institute for Technical Physics and Materials Science, Centre for Energy Research,  Hungarian Academy of Sciences, 29-33 Konkoly Thege Miklós st., H-1121, Budapest, Hungary

3       Research Group of Proteomics, Institute of Biology, Faculty of Science, Eötvös Loránd University, 1/C Pázmány P. Walkway, H-1117, Budapest, Hungary

4       Plósz Microengineering Office Ltd., 119 Üllői st., H-1091, Budapest, Hungary

5       MTA EK NAP B Research Group for Implantable Microsystems, 29-33 Konkoly Thege Miklós st., H-1121, Budapest, Hungary

**E-mail:** marton.gergely@ttk.mta.hu

**Abstract**

*Objective*. Exploring neural activity behind synchronization and time locking in brain circuits is one of the most important tasks in neuroscience. Our goal was to design and characterize a microelectrode array (MEA) system specifically for obtaining in vivo extracellular recordings from three deep-brain areas of freely moving rats, simultaneously. The target areas, the deep mesencephalic reticular-, pedunculopontine tegmental- and pontine reticular nuclei are related to the regulation of sleep-wake cycles.

*Approach*. The three targeted nuclei are collinear, therefore a single-shank MEA was designed in order to contact them. The silicon-based device was equipped with 3×4 recording sites, located according to the geometry of the brain regions. Furthermore, a microdrive was developed to allow fine actuation and post-implantation relocation of the probe. The probe was attached to a rigid printed circuit board, which was fastened to the microdrive. A flexible cable was designed in order to provide not only electronic connection between the probe and the amplifier system, but sufficient freedom for the movements of the probe as well.

*Main results*. The microdrive was stable enough to allow precise electrode targeting into the tissue via a single track. The microelectrodes on the probe were suitable for recording neural activity from the three targeted brainstem areas.

*Significance*. The system offers a robust solution to provide long-term interface between an array of precisely defined microelectrodes and deep-brain areas of a behaving rodent. The microdrive allowed us to fine-tune the probe location and easily scan through the regions of interest.

## 1. Introduction

Recording electrical activity in different brain areas simultaneously is increasingly attracting the attention of the neuroscience community [1-3], which creates new approaches in the system design of implantable devices as well. One of the main aims of researchers is to record as many neurons as possible from different brain areas, searching for synchrony and time-locking of activities. The applicable number of electrodes is limited by the tissue injury made by a certain electrode construction and the size and distribution of recording sites. Up to now, plenty of different electrode designs have been introduced such as multiple microwire arrays or bundles [4-7], micromachined silicon-based microelectrode arrays (MEAs) [8], polymer- [9], ceramic-based [10] and titanium [11] probes. When extracellular electrodes are chronically used, signal detection is hindered by glial scar formation around the implants [12, 13]. This is one of the reasons why it is advantageous if the implanted probes can be precisely moved postoperatively. This function can be enabled with the use of microdrives [14, 15].

Modern silicon-based microtechnology, fuelled by the electronic industry allows precise batch fabrication of microelectromechanical systems (MEMS) with reproducible features and small dimensions, including versatile and cost-effective ways of making neural MEAs [16]. Numerous electrodes can be placed at different points along each probe, moreover, they offer the potential of integrating signal processing circuits on them. Silicon, along with the usually applied silicon dioxide and silicon nitride insulator thin films [17], is a biocompatible material [18, 19]. Silicon can be tailored e.g. with deep reactive ion etching (DRIE) in order to form versatile probe geometries [20]. Thin-film microelectrodes, located on the silicon probes are suitable for in vivo local field potential and unit recordings [21-23]. They are widely used in freely moving animals for measurements from different brain regions, such as the neocortex or the hippocampus [24, 25]. Silicon-based microtechnology allows one to equip neural probes with delivery channels for injecting chemical compounds close to the active recording sites [26-28] or optical waveguides for optogenetic experiments [29]. According to these considerations, silicon is an excellent candidate to be substrate material for neural MEAs. However, mechanical properties of silicon microstructures are not always satisfactory. Not only the shafts of the probes can break, but cracking and degradation of the electric wires can occur. Such failures were detected in 15 µm thick MEAs, chronically implanted into mice [30]. Failures due to poor mechanical properties might be the main reason why silicon-based MEAs are hardly used in deep-brain regions even in studies involving rodents (with some exceptions [31]), while mechanical strains can be larger in case of higher mammals [30]. Recent studies revealed that silicon probes with "ultra-long" shafts can be successfully fabricated [32, 33], and these reinforced or relatively thick devices might be able to withstand larger forces [34]. The wide-spread application of ultra-long" silicon-based MEAs for discovering deep-brain regions might contribute to the progress of neurosciences.

Our neuroscientific aim was the simultaneous recording of neural activity from three nuclei of the brainstem reticular formation: the deep mesencephalic reticular nucleus (DpMe), the pedunculopontine tegmental nucleus (PPTg) and the oral part of the pontine reticular nucleus (PnO).

The targeted structures are part of the ascending reticular activating system (ARAS) and play a key role in the regulation of different functional brain states such as the sleep-wake cycle, hypersynchronization, desynchronization and arousal through integrating and transmitting different sensory inputs from the spinal cord and information from other brainstem areas [35, 36]. The ARAS consists of neuron-poor nuclei and a gigantocellular neural networks, called the reticular formation. Although these nuclei seem separate, many of them are overlapping and they are technically one anatomical unit. These areas give ascending projections to the cortex, the basal ganglia, and the thalamus, ascending descending connections to the spinal chord, and they are richly interconnected with each other [37]. Selective lesion of the PnO induces chronic comatic state, thus the PnO likely acts as an activating centre [36].

Because of the rich anatomical interconnections [38, 39], investigating functional correlation of neuronal activity from these three nuclei is an intriguing question. Extracellular neural activity of the selected areas has separately been measured in head-restrained or anaesthetized animals [40-44]. Nuclei of the ascending reticular activating system were targeted by a number of experiments using microwires in freely moving cats and rats [43, 45, 46]. However, these results are restricted to one area at a time. To the authors' knowledge, no simultaneous extracellular unit recordings from several reticular formation areas in freely moving rats have been presented. Insulated wire electrodes are frequently employed for recordings from other deep brain areas as well [47].

In contrast with these techniques, we have decided to create a probe utilizing the above described silicon-based MEMS technology, in order to interface with the deep brain structures in our interest. We expected that the silicon shaft would allow sufficient, more precise targeting than microwires, moreover, by implanting a specific formation of custom-designed array of electrodes at once, instead of managing individual wire electrodes or bundles of them, time and effort could be saved. The brain structures of interest are collinear, so we decided to create a single-shaft MEA with electrodes custom-arranged into three groups. We also aimed to construct a microdrive in order to finely relocate the sensor and find spots for successful electrophysiological recordings, containing unit activities. Another trivial way of interfacing with the three nuclei would have been the utilization three individual MEAs. Mounting a microdrive system on the skull of a rat, which is suitable for handling the shafts of three probes is probably a challenging, yet not impossible task. That would have allowed more freedom in positioning, but caused more tissue damage and cost more preparation time and resources.

In summary, this work focuses on a system, with the help of which simultaneous recording of unit activities of three deep brain structures was successfully performed in behaving rats for the first time. It will describe the measurement methods, the neurobiological analysis of the obtained results exceeds its scope.

## 2. Materials and methods

### 2.1. Probe design

A series of masks have been designed to fit the electrode array geometry to the anatomy of the brain regions we proposed to investigate. Accordingly, the total length of the probe is 30 mm, the shaft length is 25 mm, and the shaft thickness is the 200 μm initial thickness of the Si-wafer. The width of the probe gradually increases from 60 μm up to 130 μm along the recording sites, and up to 400 μm at the base of the probe to keep the robustness of the design. The distance between the tip and the middle of the first recording site is 92 μm. The tip angle is 19°. This design contains square-shaped (12 μm x 12 μm) platinum recording electrodes in three groups, located up to 3.2 mm from the probe tip. One group consists of 4 electrodes with 150 μm site spacing, and the groups are 780 μm distant from each other. The output platinum leads are 4 μm wide and 300 nm thick.

### 2.2. Microfabrication

The fabrication technology of the multielectrodes is based on a single-side, three-mask bulk micromachining process performed in three phases: (a) bottom passivation layer deposition and lift-off process for the formation of platinum electrodes, output leads and bonding pads, (b) top passivation layer deposition and patterning, (c) etching of the passivation layers and (d) deep reactive ion etching (DRIE) of silicon to form the probe body. The overall process flow is summarized in figure 1.

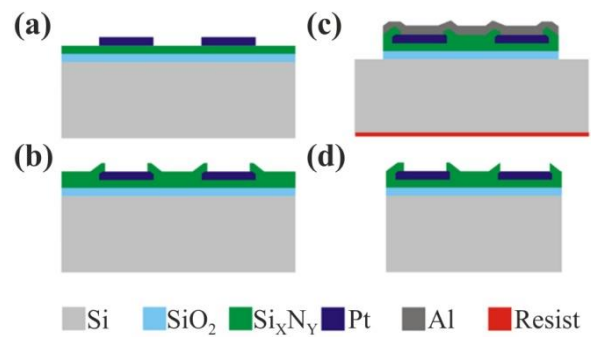

Si    SiO₂    SiₓNᵧ    Pt    Al    Resist

**Figure 1.** Cross-sectional schematic diagram of the main fabrication steps. (a) $SiO_2$ and $Si_xN_y$ layer deposition and patterning of platinum wiring. (b) Deposition and selective opening of the $Si_xN_y$ passivation layer. (c) Patterning of front-side Al mask and dry etching of the passivation and insulation layers; evaporation of Al and spin-coating of photoresist to form the backside etch-stop layer. (d) Cross-section of the ready-made probe after removing the masking and etch-stop layers.

### 2.2.1. Electrode array formation

200 μm thick, double-side polished (100) oriented 4-inch silicon wafers are used for probe fabrication. In the first step a 100 nm thick $SiO_2$ layer was thermally grown, followed by deposition of a 500 nm thick LPCVD low-stress silicon nitride. The metal layer was then deposited and patterned by lift-off process. The lift-off structure consisted of 1.8 μm thick photoresist (Microposit 1818) layer over patterned 500 nm thick Al thin film. The metal layer consisted of a 15 nm thick adhesion layer of $TiO_x$ and Pt. $TiO_x$ was formed by reactive sputtering of Ti in $O_2$ (Ar/$O_2$ ratio was 80:20) atmosphere. In the same vacuum cycle 270 nm thick Pt was sputtered on top of $TiO_x$. The lift-off was accomplished by dissolving the photoresist pattern in acetone, this process was optimized by using water-cooled substrate holder which diminished the resist distortion during $TiO_x$/Pt sputtering. Subsequently, the removal of Al patterns in a fourcomponent etching solution and low pressure chemical vapor deposition of 300 nm thick $SiN_x$ insulating layer stack were performed. Contact holes were etched through the $SiN_x$ layer to unblock the electrical recording sites at the probe tip and the pads at the base for wire bonding.

### 2.2.2 Probe shaping

A 500 nm thick Al layer was deposited on the front side and patterned by photolithography using 4 μm thick SPR220 photoresist. The contour of the probe body was defined by Al wet etching. SPR220 photoresist was spun on the backside of the wafer acting as a stopping layer during the subsequent deep reactive ion etching of silicon. The 3D micromachining process was performed in an Oxford Plasmalab System 100 chamber using Bosch process.

### 2.3. Packaging

Rigid-flexible PCBs have been formed in order to provide sufficient support for the silicon probes and enough freedom for the movements of the microdrive. The deformability of this component is illustrated in figure 2. The substrate of its rigid parts was made of 1-mm thick FR-4. The substrate of the flexible cable was manufactured from 50 μm thick polyimide. A conductor layer of copper has been formed on the top surface, plated with gold for the ease of wire bonding. The upper insulator layer on the PCB was made of flexible solder-resist mask. The geometry of the device has been designed as follows. A 6 mm x 9 mm rigid part was shaped for the support of the probe. A hole of 2 mm in diameter was designed for fixing the PCB with a screw onto microdrive. The L-shaped flexible cable was 5 mm wide and 19.1 mm long at its midline. It provided electronic connection via 200-μm thick wires between the probe support and the Preci-Dip connector part of the PCB. The connector part was rigid, suitable for soldering a 12-pin section of a 90-degree angle pin connector (Preci-Dip 850-PP-N-050-20-001101).

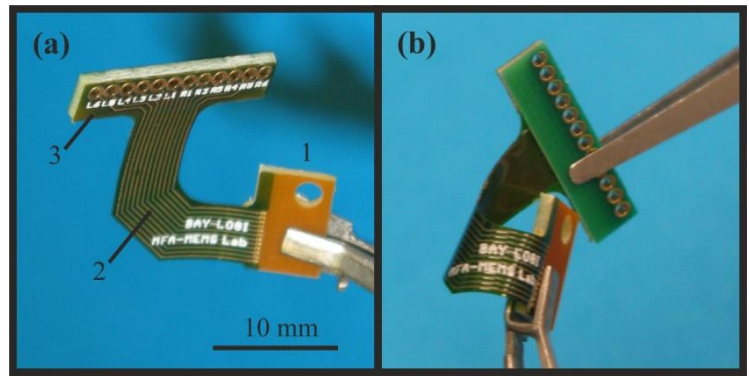

**Figure 2.** The rigid-flexible PCB, containing a rigid part for probe support (1), the flexible polyimide cable (2) and the Preci-Dip connector part (3).

### 2.4. Microdrive

The extended length of the probe requires stable and sufficiently vibration-free system to minimize tissue damage along the probe track, while providing flexibility to move the implanted electrode shank in the brain. A novel microdrive configuration was designed and is illustrated in figure 3. The 8 mm wide, 7 mm thick, 25 mm high, body of the device weights 2.7 g and consists of three, half octagonal prism-shaped blocks. The upper block (5), which contains the thread of the driving screw (5), and the lower block (7) are made of aluminium. The electrode (4) can be fastened through the rigid part of the PCB (3) to the microdrive by a small screw (9). The middle, electrode-carrier block (6) is held by two tightly fit rods (8) and is driven by the large driving/positioning screw (10). This block is made of polytetrafluorethylene (PTFE), as this self-lubricating type plastic renders the opportunity to highly reduce the joint gap between the block and the rods. Thus vibration is reduced to the minimum, resulting in a really robust and stable system. One 360° turn of the driving screw moves the probe tip with 300 μm, without any rotational movement.

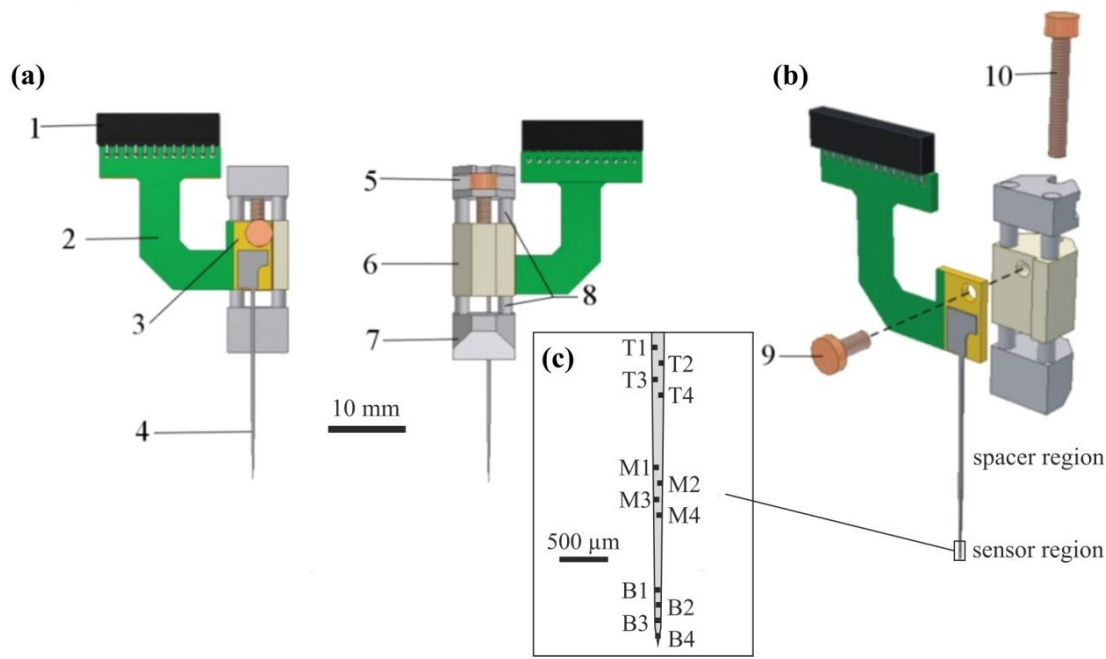

**Figure 3.** Layout of the Si-electrode - microdrive system. (a) Front (left) and rear (right) view of the assembled system. 1: connector; 2: flexible cable; 3: the rigid part of the PCB; 4: the probe shaft; 5: microdrive upper block with the thread for the driving screw; 6: microdrive middle, electrode-carrier block; 7: microdrive lower block; 8: two tightly fit rods; 9: fastening screw; 10: driving screw. (b) Assembling of the driving screw, the microdrive and the electrode via the fastening screw. (c) Schematic drawing of the sensor region of the shaft, with three groups of sites.

*2.5. Electrochemical impedance spectroscopy*

In order to characterize the electrodes, electrochemical impedance spectroscopy (EIS) measurements were performed in physiological saline, using an Ag/AgCl reference electrode (Radelkis Ltd., Hungary) and a counter electrode of platinum wire with relatively high surface area. The probe signal was sinusoidal, 25 mV RMS. A Reference 600 instrument (Gamry Instruments, PA, USA) was used as potentiostat and Gamry Framework 6.02 and Echem Analyst 6.02 software were used for experimental control, data collection and analysis. Experiments were performed in a Faraday cage.

*2.6. Animal surgery*

Eleven adult Sprague-Dawley rats (Charles River Laboratories, from the breeding colony of ELU, Hungary) were implanted with the probes under 4% isoflurane anaesthesia. The care and treatment of all animals were conformed to Council Directive 86/609/EEC, the Hungarian Act of Animal Care and Experimentation (1998, XXVIII), and local regulations for the care and use of animals in research. All efforts were taken to reduce the number of animals used and to minimize the animals' pain and suffering. Prior to surgery we chronically applied the antibiotic Enorofloxacin in 10 mg/kg dose to prevent inflammatory reactions. We mounted the animals onto a stereotaxic instrument (David Kopf Instruments, California, USA), and a longitudinal cut was made on the scalp. Skin and muscles were retracted and the surface was cleaned. Four screw electrodes over the cortex were used to record the electrocorticogram (ECoG): one over the right secondary motor cortex (AP: 4.5 mm, ML: 2.5 mm with reference to the bregma), one over the right somatosensory cortex (AP: -2.9 mm, ML: 2.5 mm), two over the right and left visual cortex (AP: -8,3 mm, ML: 2,5 mm). Two additional screw electrodes over the cerebellum served as grounding. After driving the screw electrodes into the cranium, the skull was opened over the left hemisphere using a drill, and the dura mater was incised. The neural probe was fastened to the microdrive that was adjusted to the stereotaxic instrument through a custom-made adaptor, and the probe was inserted into the brain manually by oblique targeting (47° inclination to the vertical), which made it possible to span all three target areas through a single electrode track at the same time. After the initial penetration, silicone-vacuum grease (Beckman Coulter Inc., Fullerton, CA, USA) was applied on the probe shank at the brain surface to provide smooth penetration further into the tissue and prevent blood clot formation. Finally, SuperBond resin cement (Sun Medical Company Ltd., Moriyama, Japan) was applied on the skull surface and the microdrive was fixed. A crown-like structure was formed around the system, utilizing UniFast dentacrylic cement (GC America Inc., Chicago, USA) and sterile plastic components fabricated from syringes and Petri dishes, protecting its external components. The structure typically weighted 12 g, was 28 mm long, 22 mm wide, and 30 mm high. The microdrive – and thus the implant - was not sealed hermetically from the environment, therefore proper hygienic conditions were in focus to avoid infection at the craniotomy. The implantation is illustrated in figure 4.

## 2.7. Course of experiments

During recordings, the rat was kept in a cage of 40 cm × 60 cm, surrounded by a Faraday cage for electrical shielding. The connecting cable was continuously held up by a light mechanical lever, which prevented the rat from reaching the cable and damaging it, without impeding the animal in its movements.

Nine rats were allocated for gaining high amount of data considering the characteristics of the targeted brain nuclei. To achieve this, we needed short, 1-2 hour sessions, during which the rats changed from awakeness to sleeping state or vice versa. During these sessions, we measured firing patterns of individual cells. After a successful session (or on the following day) we used the microdrive to relocate the probe so that new groups of cells could be explored. During the initial implantation, the probe tip was aimed at the area of the pedunculopontine tegmental nucleus. The Bregma was used as reference point for targeting, Instead of surface coordinates, a depth coordinate was determined (AP -7.0 ; ML 1.5; dorsoventral 7.6), to which the tip was targeted in a 47° angle and from where the oblique path down to deeper areas started. The tilting of the head in the sagittal plane was avoided by maintaining the "flat skull position" by checking and refining the correctness of the lambda (DV = 0) and interaural line (AP = 9.0; DV = 10.1) coordinates in reference to the bregma. During insertion, the probe was slowly (approximately 2-3 mm/min), manually advanced, solely with the usually dorsovenrtral directional drive of the stereotaxic apparatus (which was tilted so that the driving direction was parallel to the direction of the probe shaft).

Thus each electrode group could record from different deep brain areas over time. At the deepest penetration the tip reached the caudal pontine reticular nucleus. Altogether, the system allowed us to record 12 unit channels (4 from each target area) and 4 ECoG channels from each animal. The electrode array was slowly advanced (60-100 μm per day) during the recording period (minimum 22, maximum 125 days) and stopped if extracellular unit firing coud be recorded, until the eledote tip reached the PnC.

Two additional rats were sacrificed in order to gain some information about the chronic static performance of the electrodes (without driving). In this case, we used the very same methodology for implantation, as previously described, but did not drive the probe so frequently, as instead, we kept them at the same location for weeks.

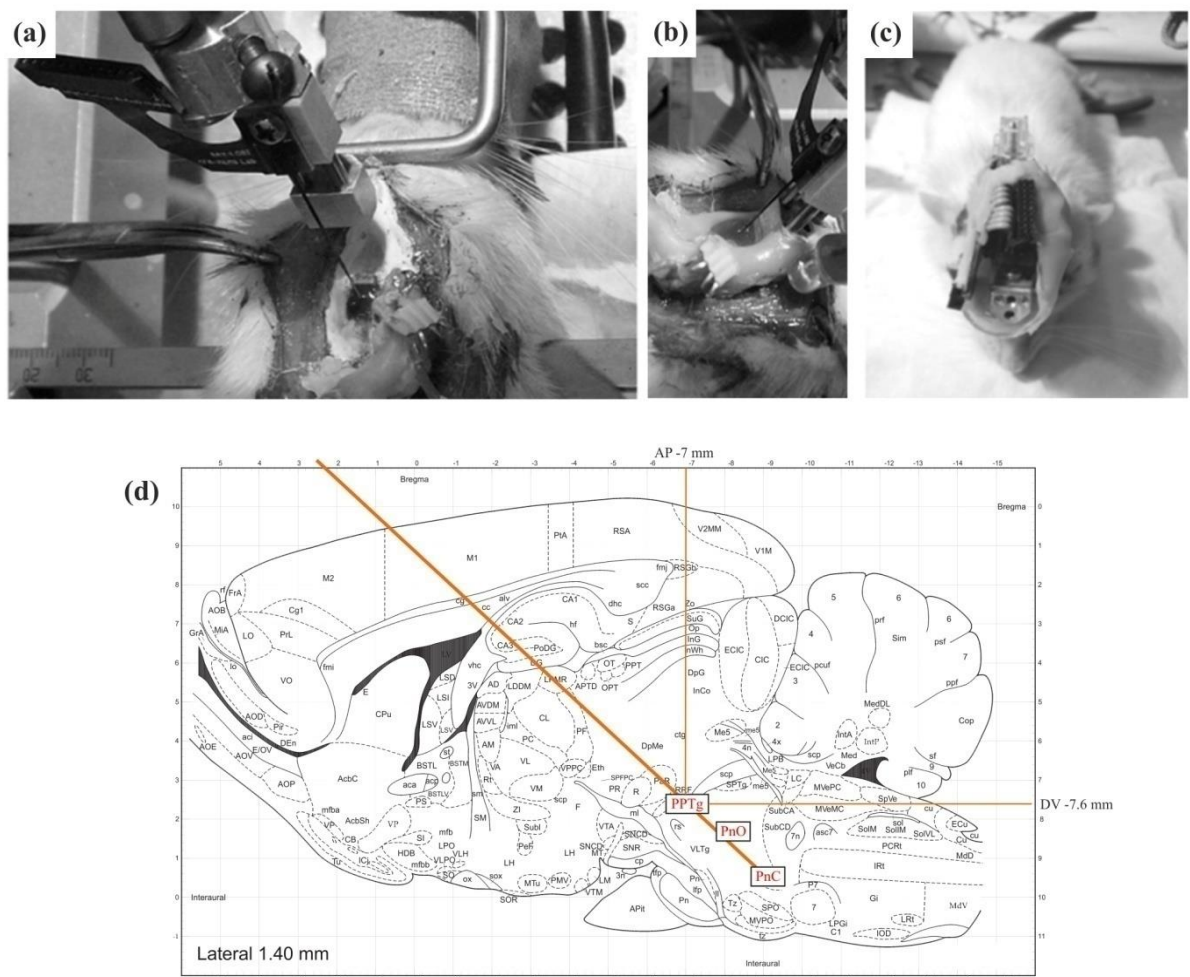

**Figure 4.** Implantation surgery and histological evaluation of electrode position. (a), (b) Positioning and inserting of the electrode into the brain tissue. (c) Implanted rat after surgery. (d) Illustration of the implantation strategy. Figure based on Fig. 82. of [48]. PnC: pontine reticular nucleus, caudal part; PnO: pontine reticular nucleus, oral part; PPTg: pedunculopontine tegmental nucleus

*2.8. Recording system and data analysis methods*

We used a multichannel amplifier (Multiamp SMA-1a, Supertech) and a data processing system (CED Micro1401, Cambridge Electronics Design Ltd., Cambridge, UK) for data acquisition. Extracellular action potentials were captured at 33 kHz on each electrode whenever their amplitude reached the 100 μV threshold. We used a 500 – 8000 Hz online filter and a 300 – 6000 Hz offline filter before spike sorting. Electrocorticogram digitalization frequency was 5 kHz (offline filtering between 0.2 and 300 Hz). Data files were recorded with Spike2 (Cambridge Electronic Design Ltd., Cambridge, UK) software. Spike waveforms were sorted by using a 3D off-line spike analysis program (OFS/v2, Plexon Inc., Dallas, TX, USA). With the help of principal component analysis, we determined scatterplots using the first two principal components. K-means clustering analysis was used to define cluster borders of single units. ECoG spectrograms, autocorrelations and cross-correlations were calculated to validate the cluster verification with NeuroExplorer (Plexon Inc., Dallas, TX, USA). To assess the quality of spike cluster separation, we investigated the spike contamination of the refractory period in the autocorrelograms [49]. If such contaminations were present, additional efforts, such as manual clustering were made to separate the waveforms.

In order to quantify the quality of the unit cluster separation, we employed three statistical methods: a multivariate analysis of variance (MANOVA) test (F), a nonparametric test ($J_3$) and the calculation of the Davies-Bouldin validity index (DB), which is also nonparametric. These methods are described in detail in the supporting text of a research paper by Nicolelis et al. [50]. The $J_3$ index is calculated by dividing the average distance between clusters ($J_2$) by the average distance of points from their cluster mean ($J_1$). Therefore, higher $J_3$ values indicate better separation rates. The DB index is defined as

$$DB = \frac{1}{n}\sum_{k=1}^{n} \max_{k \neq l}\left\{\frac{S_n(Q_k)+S_n(Q_l)}{d(Q_{k,}Q_l)}\right\},\qquad(1)$$

where n is the number of clusters, $S_n(Q)$ is the average distances of the elements of cluster Q to the center of cluster Q and $d(Q_k,Q_l)$ gives the distance between the center of the $Q_k$ and $Q_l$ clusters [50]. Good clustering is characterized by high F, J3 and small BD index values.

*2.9. Histological evaluation*

Electrode placement can be verified by electrical stimulation through the recording sites in rats deeply anesthetized with urethane [51, 52]. We used 200–400 ms square wave current pulses with a current strength of 1–10 µA. Animals were then transcardially perfused with 4% paraformaldehyde in sodium-phosphate buffer. 60 µm longitudinal sections were cut from the brains using a freezing microtome, and the tissue was stained with the Gallyas silver-staining method [53, 54]. Thus the electrode track of the chronically inserted probe and acutely injured „dark" neurons caused by the electrical stimulation can be seen at the same time.

## 3. Results and discussion

*3.1. The realized system*

The custom design technology allowed precise electrode array arrangement on the silicon probe, fitted to the location of the three brain areas we intended to observe. Photographs and scanning electron microscopic images of the MEA are shown in figure 5(a)-(c). The results of the EIS measurements are shown in figure 5(d). At 1 kHz, 3.12 MΩ ± 0.21 MΩ average impedance magnitude was yielded for the $12 \times 12$ µm² electrodes, which is in consistence with the literature [55]. The robustness of the device has proven to be sufficient for implantation. We have observed neither breaking nor bending of the shafts. The flexible connection to the socket and firm holding of the electrode in a hard printed circuit board (PCB) attached to the microdrive kept the stability of the probe and provided freedom for driving.

While custom designed microdrives are commonly used for the manipulation of fine wire electrodes, tetrodes or hexatrodes [56-58], it is less frequent in case of silicon-based MEAs. With a silicon-based MEMS thermal actuator, driving of a microelectrode array with 8.8 µm step resolution was achieved [59]. However, the construction of the device requires deposition of 11 layers onto the substrate wafer, including 4 layers of polysilicon. In our case, the driving was achieved without involving subtle motors or active elements. Compared to the dDrive system of NeuroNexus (Ann Arbor, MI, USA), the system presented here utilizes aluminium blocks instead of plastic, and two vertical aligners improve stability of the vertically moving block rather than purely relying on the driving screw. The whole system met the particular demands of the angled implantation and deep location of the targeted areas.

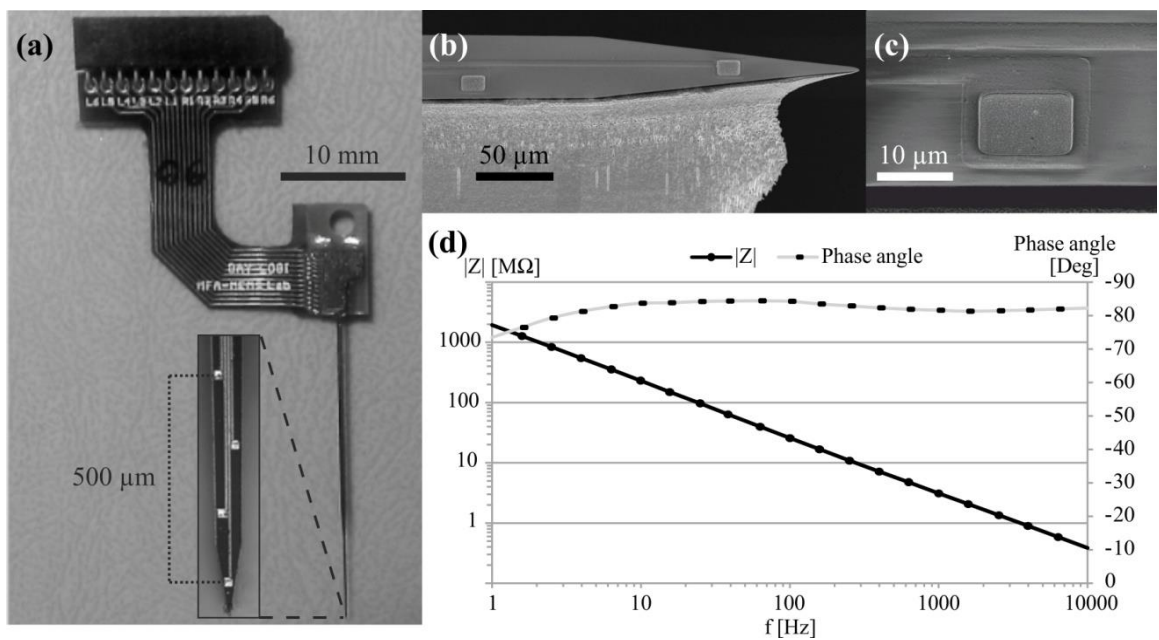

**Figure 5.** (a) The silicon probe mounted onto the rigid-flexible PCB. (b)-(c). Scanning electron microscopic images of the probe tip and a recording site. (d) Average EIS spectrum of the electrodes of a probe. Standard deviation values were less than 10% of the corresponding magnitude values on all frequencies. At 1 kHz, the magnitude of the impedance was 3.12 MΩ ± 0.21 MΩ.

The microdrive was stable during the recording sessions. The stereotaxic instrument provided a specific targeting angle and precise implantation during surgery, and the microdrive allowed uni-directional positioning of the MEA even after six months. Figure 6 shows images of the histological findings. The dark neurons of the Gallyas staining method indicated an approximately 4-600 µm wide trail of perturbed tissue along the probe track and confirmed successful targeting of the deep-brain regions of interest. No sign of bending was observed on the used, explanted probes.

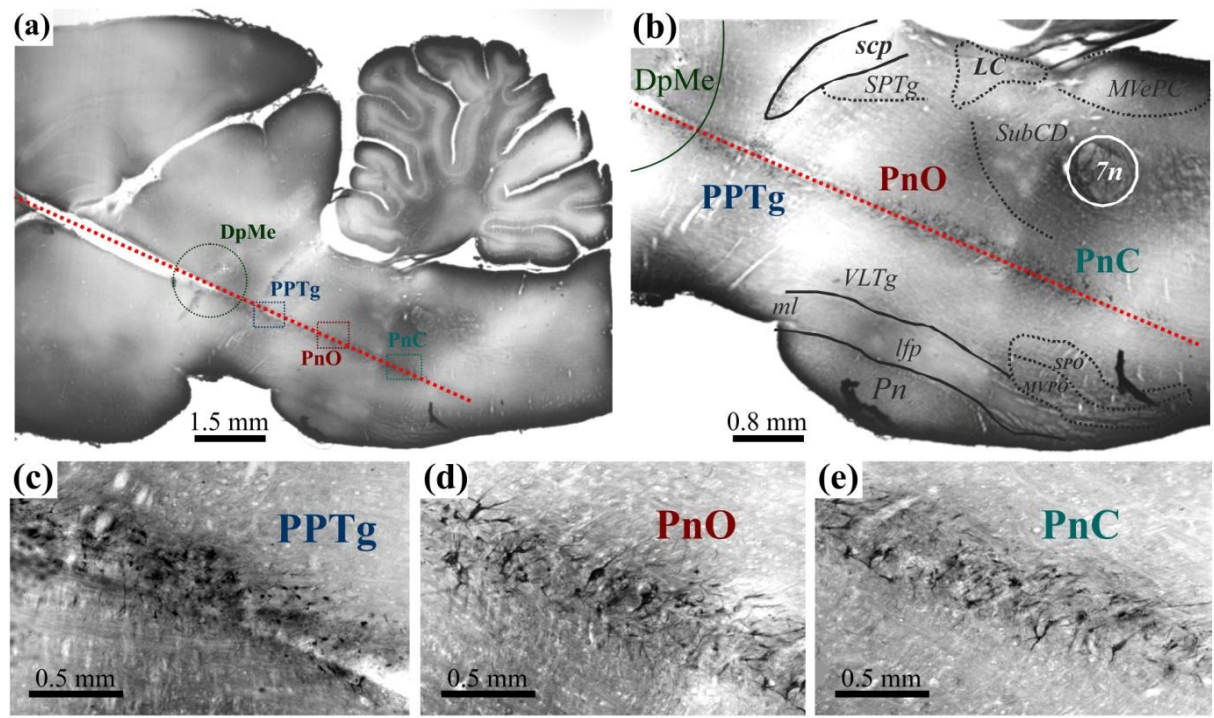

**Figure 6.** Electrode position evaluated with histology. Slices were stained with Gallyas method. Dark neurons indicate the position of the probe shaft in the brain tissue. (a)-(b) Electrode track with a schematic drawing of the inserted probe. Scale bars are respectively 1.5 mm and 0.8 mm. Abbreviations: 7n: facial nerve or its root; DpMe: deep mesencephalic nucleus; LC: locus coeruleus; lfp: longitudinal fasciculus of the pons; MVePC: medial vestibular nucleus, parvicellular part; MVPO: medioventral periolivary nucleus; Pn: pontine nuclei; PnC: pontine reticular nucleus, caudal part; PnO: pontine reticular nucleus, oral part; (c)-(e) Zoomed images of the probe track in the targeted nuclei. Scale bars are 400 μm.

### 3.2. Unit activities recorded from the targeted nuclei

The majority of the implanted MEAs (n = 9) were utilized in order to obtain single unit activities from different cells in the nuclei of freely moving rats. As mentioned earlier, in section *2.7. Course of experiments*, these probes were driven further with the microdrive after successful, 1-2 hour recording sessions, which contained transitions from awake/sleeping states. We have successfully recorded single unit activity chronically from all target areas (DpMe, PpTg, PnO). Figure 7 shows examples of recorded units during an epoch of 15 seconds in a rat, 25 days after surgery. Raw data and spectrogram of the ECoG, raw data and unit firing from 4 multi-unit recording sites and the averaged waveforms of sorted individual units are presented. The spectrogram of the ECoG reveal sleep spindles, starting at around t = 2 and 14 seconds. Spike trains of cells are presented that fire frequently at the beginning of the spindle (M1 units), and ones that cease to be active at the end of the spindle (B2, B3 units). The data gained by the MEAs gave insight into the firing patterns of different cells that take part in sleep regulation in the brainstem.

We were able to record approximately 200 extracellular units from this animal, of which 163 were thoroughly analyzed after sorting. Units that fired for just a short period of time or had too low firing frequency or high spike waveform variance were excluded. Of the 163 neurons, 15 were present on a single electrode site only. Thus we recorded 148 neurons concurrently. The active contacts between parallel recordings were minimally 2, maximally 6 at the same time. The contacts showed unit activity in ~27.5% of the cases with an average of 5 recorded cells. These results show that individual cell activities in deep brain regions of a rodent can be recorded with relatively high yield using such silicon probes.

The number of simultaneous recordings of sorted units from the target nuclei (DpMe, PPTg, PnO) was achieved, however, sparsely: altogether 13 such sessions occurred in the 9 rats out of altogether 266 analyzed sessions (on the average, 1.4 ± 0.5 per rat).

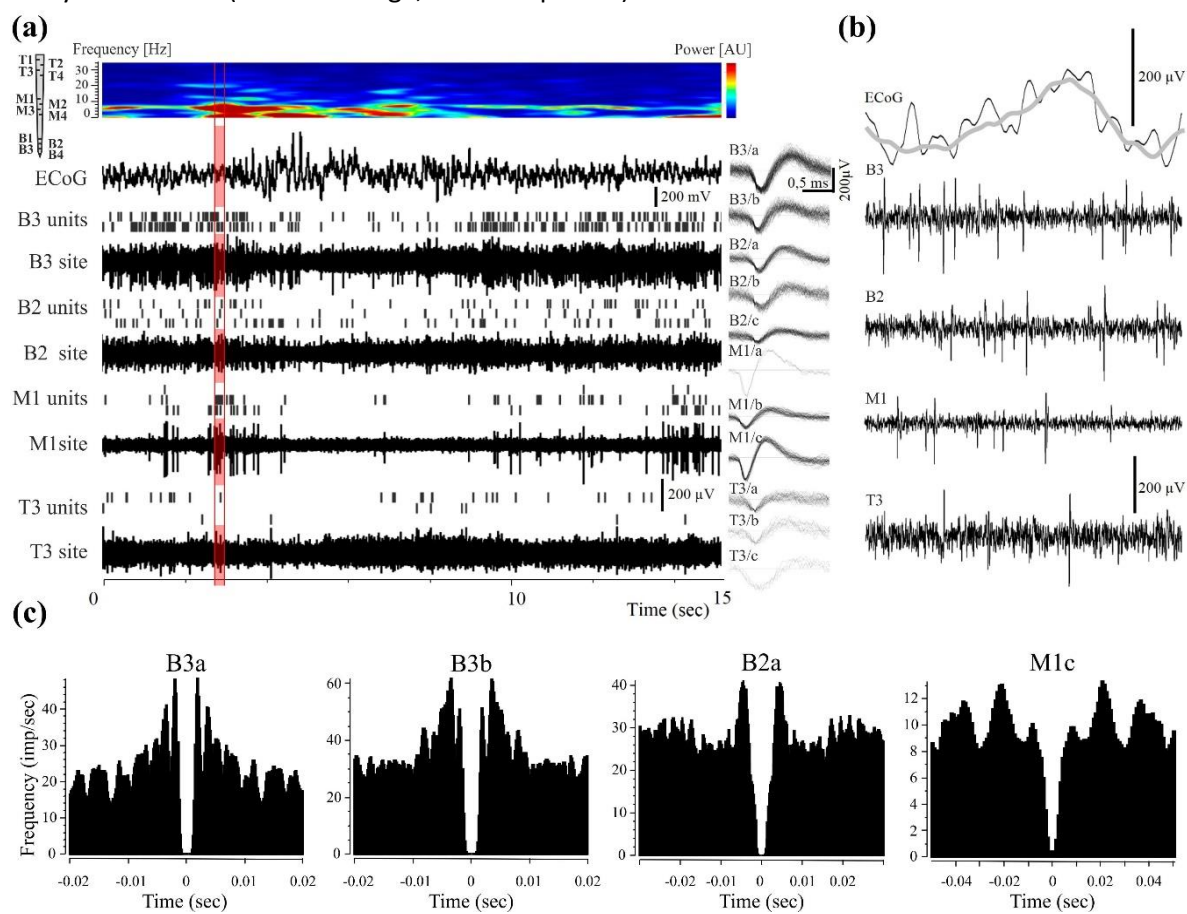

**Figure 7.** Representative parallel recordings from different sites on an implanted probe in a freely moving rat, 25 days after surgery. (a) Raw data and spectrogram of the ECoG channel, and unit channels with single unit patterns (spike trains). Elevated power in the lower frequency range of the ECoG channel indicates sleep spindles starting at around t = 2 and 14 seconds. Activities of cells that are in connection with the spindles were recorded. (b) 0.24 second long recording sections (highlighted with a red band on panel A) at a faster time-base after offline filtering (ECoG: 0.2 – 300 Hz, probe site channels: 300-6000 Hz). (c) Different types of representative autocorrelograms show heterogeneous spiking properties.

*3.3. Signal stability*

Two additional rats were implanted in order to gain insight into the static chronic performance of the electrodes. These MEAs were not driven after each successful recording session, as instead, we let them stay in the same location for 2-7 days. Figure 8 shows the functionality of such a probe during a 33-day period. The number of different single unit activities varied from 1 to 8, with an average of $3.9 \pm 2.0$. Figure 8(a) shows the number of single unit activities on each day, when recordings were performed. Black triangles mark the occasions when the probe was driven further into the tissue. The quality of unit clustering is shown in figure 8(b). The mean $\pm$ standard error for the indices throughout the 33-day period were the following: $F = 8.23 \pm 5.43$, $J_3 = 2.76 \pm 1.53$, $DB = 0.26 \pm 0.1$. For these values, the measurements of Nicolelis et al. with microwire arrays in the cortex of rhesus monkeys yielded $F = 11.5 \pm 1.0$, $J_3 = 3.8 \pm 0.4$, $DB = 0.32 \pm 0.01$ [50]. Figure 8(c)-(c') illustrate the results of cluster analysis of three cells recorded from the same animal on day 21 and 22, respectively, showing that good quality single units of the same clusters can be recorded on different days.

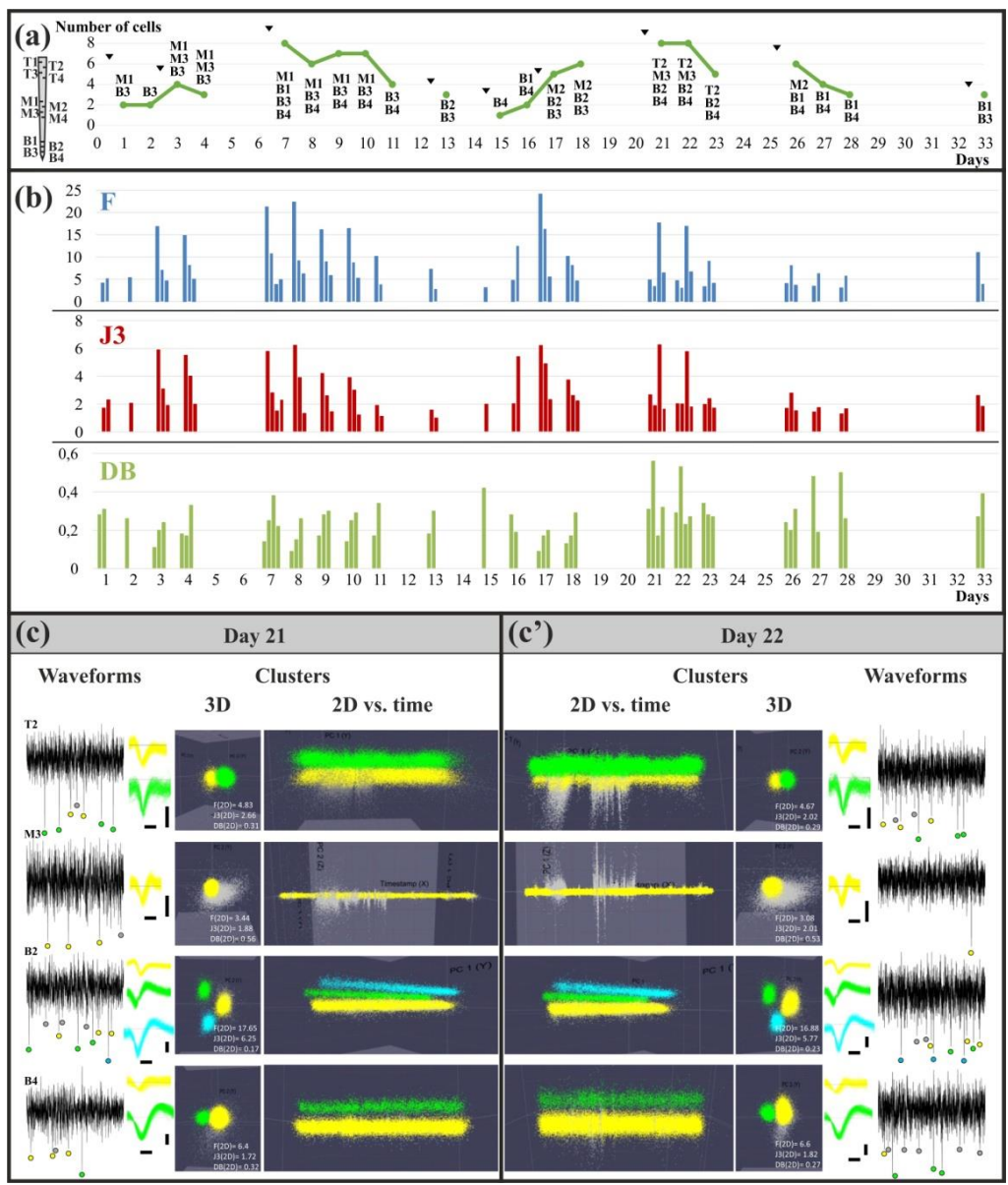

**Figure 8.** (a) The number of single unit activities on each day, when recordings were performed from a rat. Black triangles mark the occasions when the probe was driven further into the tissue. (b) The quality of unit clustering. The mean ± standard error values considering the 33-day period: F = 8.23 ± 5.43, $J_3$ = 2.76 ± 1.53, DB = 0.26 ± 0.1. (c)-(c') Typical unit clusters on a channel during a session and the same clusters obtained from the signals recorded on the following day. Horizontal and vertical scale bars on the spike waveforms are 0.5 ms and 100 µV, respectively. The cluster plots contain data of 120 minute long recordings.

Unit activities on 20-30% of the electrodes on 15 µm thick planar Michigan MEAs were reported 30 days post-implant, in the motor cortex of rats [60]. With such probes, 10-15% unit yield was measured after 3-6 months, even though a significant portion of the electrodes was in white matter [13]. Comparing our results to those findings, the long-term performance of our system is slightly poorer. The causes of this can be various, we provide three possibilities here. Firstly, highly different brain regions might respond differently to the implantation: the cells in our targeted nuclei might be more sensitive compared to the cells of the cortex or the hippocampus. Such heterogeneity in unit stability was even observed in different layers of the cortex: layers IV to VI were claimed to have the greatest yield acutely, but layers II to IV the best yield in chronic time points [13]. Secondly, the much larger (200 μm) shaft thickness of our probes is disadvantageous in this aspect, it can induce a more aggressive immune response [61]. Finally, interfacing with the deep brain nuclei required a much longer implantation track than the cortical areas would have required, which results in a bigger chance of disrupting blood vessels, thus facilitating chronic signal degradation [12].

It was not our goal to measure the same unit activities for weeks or months (which is a common priority e.g. for the development of brain-computer interfaces with invasive electrodes). Rather, we intended to observe firing patterns of different cells of deep brain nuclei during 1-2 hour sessions. In the light of our aims, the microdrive alleviated the problem of losing signals of unit activities, as slight motions of the probe could bring the electrodes into less perturbed tissue regions and explore new cells.

## 4. Conclusions

We report here on a novel silicon-based MEA design, tailored to be used in a specific neuroscientific experiment, which required interfacing with three deep brain nuclei related to the regulation of sleep-awake cycles. The applied fabrication technology provided cost-effective batch production of the MEAs. The stiff body of the probe allowed sufficient implantation, even into the brainstem, while the custom-designed electrode array formation provided coverage of the investigated brain regions. The probe was suitable for measuring unit activities in the targeted structures, with high yield, through several recording sites of freely moving rats, fulfilling our goals. The main novelty of our system is the utilization of a special microdrive, which enabled us to accurately relocate the electrodes after the implantation, and explore new cells. Taking into account that the brainstem is one of the most difficult targets for neural recording in rats, our system can probably be used in many different experiments, when two or more brain structures can be exposed in an electrode track, even if the track is not perpendicular to the cerebral surface. In a future study, we intend to investigate the usability of such MEAs in the central nervous system of larger mammals, such as cats or primates.

## Acknowledgements

We thank András Czurkó for critical reading of the manuscript. We are thankful for the laboratory personnel of the MEMS Laboratory of the Institute for Technical Physics and Materials Science. This work was supported by the KTIA_13_NAP-A-IV/1-4;6 research grants. Z. Fekete is grateful for the support of the Hungarian Brain Research Program (KTIA NAP 13-2-2014-0022). A. Pongracz is grateful for the Bolyai Janos scholarship of the Hungarian Academy of Sciences.